\documentclass[authoryear,preprint,review,3p]{elsarticle}

\usepackage{amssymb}
\usepackage{amsmath}

\journal{Statistics \& Probability Letters}

\usepackage{enumitem}

\newcommand*{\rmd}{\mathrm{d}}
\newcommand*{\KL}{\mathrm{KL}}
\newcommand*{\reals}{\mathbb{R}}
\newcommand*{\nats}{\mathbb{N}}

\newcommand*{\eqind}{\stackrel{\mathcal{D}}=}
\newcommand*{\indicator}[1]{\mathbf{1}\left\{#1\right\}}
\newcommand*{\expect}[2]{\mathbf{E}_{#1}\left[ #2\right]}
\newcommand*{\filt}{\mathbb{F}}

\newdefinition{definition}{Definition}
\newtheorem{proposition}{Proposition}
\newtheorem{theorem}{Theorem}
\newtheorem{lemma}{Lemma}
\newtheorem{corollary}{Corollary}
\newdefinition{proof}{Proof}
\newdefinition{example}{Example}

\begin{document}

\begin{frontmatter}



\title{Anytime-Valid Tests of Group Invariance through Conformal Prediction}
\author[1]{Tyron Lardy\corref{cor1}\fnref{fn1}}
\ead{t.d.lardy@math.leidenuniv.nl}

\author[2]{Muriel Felipe P\'erez-Ortiz\fnref{fn1}}
\ead{m.f.perez.ortiz@tue.nl}
\cortext[cor1]{Corresponding author}
\fntext[fn1]{Declarations of interest: none}

\affiliation[1]{organization={Mathematical Institute, Leiden University},
           addressline={Einsteinweg 55}, 
           city={Leiden},
           country={The Netherlands}}

\affiliation[2]{organization={Department of Mathematics and Computer Science, Eindhoven University of Technology},
           addressline={ PO Box 513}, 
           city={Eindhoven},
            postcode={5600 MB}, 
           country={The Netherlands}}





\begin{abstract}
We develop anytime-valid tests of invariance under the action of compact groups. 
The resulting test statistics are optimal in a logarithmic-growth sense.
We apply our method to extend recent anytime-valid tests of independence and to construct tests of normality.
\end{abstract}




\end{frontmatter}




\section{Introduction}

Suppose that we observe data $X_1,X_2,\dots$ sequentially, and that they take values in a probability space $\mathcal{X}$. 
We are interested in testing the null hypothesis of invariance of the data under the action of a sequence of groups $(G_n)_{n\in \mathbb{N}}$, that is,
\begin{equation}\label{eq:null}
    \mathcal{H}_0: gX^n\eqind X^n \quad \text{for all } g\in G_n \text{ and all } n\in \mathbb{N},
\end{equation}
where $\eqind$ signifies equality in distribution and, for each $n\in\nats$, $G_n$ is a group of transformations of the data---a compact topological group that acts continuously on $\mathcal{X}^n$. 
In order to avoid pathological situations additional structure is needed (see Section~\ref{sec:seq_action}). 
Note that the observations are not assumed to be independent nor identically distributed; they are only assumed to be sampled from a distribution on infinite sequences. 
Prominent examples of \eqref{eq:null} include testing for exchangeability \citep{vovk2005algorithmic, ramdas2022testing}, in which case $G_n$ is the group of permutations on $n$ elements; and testing for sphericity, in which case $G_n$ is the orthogonal group $\mathrm{O}(n)$.
The latter can be used to test the Gaussian-error assumption in linear regression, as we will see. We construct anytime-valid tests for $\mathcal{H}_0$ that monitor test martingales and randomized versions thereof.

A sequence of statistics of the data is a test martingale if it is nonnegative, starts at one, and is a martingale under every element of $\mathcal{H}_0$.
Formally, if $\mathbb{G}=(\mathcal{G}_n)_{n\in \mathbb{N}}$ is a filtration of $\sigma$-algebras such that $\mathcal{G}_n\subseteq \sigma(X^n)$, then a sequence of statistics $(M_n)_{n\in \mathbb{N}}$ is a test martingale for $\mathcal{H}_0$ with respect to $\mathbb{G}$ if $\expect{Q}{M_n\mid \mathcal{G}_{n-1}}= M_{n-1}$ for all $Q\in\mathcal{H}_0$ and $M_0=1$. 
Test martingales are the central objects for anytime-valid inference~\citep{ramdas2023gametheoretic}.  
A sequential test $\phi_n=\indicator{M_n \geq \frac1\alpha}$ can be built using the threshold $1/\alpha$ and the resulting test $\phi_n$  is anytime valid in the sense that it enjoys the time-uniform type-I error guarantee that $Q(\exists n\in \mathbb{N}: \phi_n=1)\leq \alpha$ for any $Q\in \mathcal{H}_0$. 
Furthermore, it will also be useful to consider randomized test martingales, for which we append an independent random number $\theta_n\sim \text{Uniform}([0,1])$ to each $X_n$.

We build test martingales using tools from conformal prediction, which is perhaps best known as a framework for uncertainty quantification for point predictors (see \citet{vovk2005algorithmic} and Section~\ref{sec:conformal_martingales}).
The test martingales constructed using these methods are known as conformal martingales. 
Most crucially for our present purposes, \citet{vovk2005algorithmic} show that conformal martingales can be used to test whether data are generated by a specific class of generating mechanisms, called online compression models. The key insight of this note is that group-invariant models can be regarded as online compression models. Therefore, conformal martingales can be built using the conformal-prediction machinery to test the hypothesis of invariance in \eqref{eq:null}.

We further show that the resulting test martingales are optimal in a specific logarithmic sense. The rationale behind this criterion is that, under the alternative distribution, a ``good'' test martingale should no longer be a martingale and it should grow large to gather evidence against the null. 
Formally, given a specific alternative $P$ of interest and a stopping time $\tau$ for the experiment, the relevant test statistic is $M_{\tau}$---the value of a test martingale $(M_n)_{n\in\nats}$ at $\tau$. For example, $\tau$ could be the first hitting time for the threshold $1 / \alpha$ as before. A common measure to judge the expected growth of $(M_n)_{n\in\nats}$ at $\tau$ is $\expect{P}{\log M_\tau}$, and test martingales that  maximize this growth rate are referred to as log-optimal~\citep[see][]{koolen2022log,ramdas2023gametheoretic}.

In Section~\ref{sec:seq_action}, we give additional structure to the problem of sequential group invariance, give the necessary background on online compression models, and we show that sequential group-invariant models define online compression models. 
Section~\ref{sec:conformal_martingales} shows our construction of test martingales using the online-compression structure and conformal prediction.  Section~\ref{sec:mart_opt} shows that our construction is log-optimal against certain alternatives. 
Section~\ref{sec:applications} applies the results to test for independence and to test rotational invariance, and
Section~\ref{sec:discussion} discusses the limitations of invariance testing in the context of online compression models and modifications of the constructions that are used.

All the proofs can be found in \ref{app:proof} in the supplementary material.

\section{Sequential group actions and online compression models}\label{sec:seq_action}
The hypothesis in~\eqref{eq:null} is only meaningful if the statements regarding group invariance for each $n$ are consistent with each other; without any further restrictions, invariance of the data at one time may contradict the invariance of the data at a later time. 
We assume that the groups $(G_n)_{n\in \mathbb{N}}$ act sequentially on the data to avoid such situations.
\begin{definition}[Sequential group action]\label{def:seq_groups}
    The action of the sequence of groups $(G_n)_{n\in\mathbb{N}}$ on $(\mathcal{X}^n)_{n\in\mathbb{N}}$ is sequential if the following conditions hold.
    \begin{enumerate}[label=(\roman*)]
    \item \label{item:ordered_groups} The sequence $(G_n)_{n\in\nats}$ is ordered by inclusion: for each $n$, there is an inclusion map $\imath_{n+1}: G_n \to G_{n + 1}$ such that $\imath_{n + 1}$ is a continuous group isomorphism between $G_n$ and its image, and the image of $G_n$ under $\imath_{n+1}$ is closed in $G_{n+1}$. 
    \item\label{item:sequential_action} For all $g_n\in G_n$ and all $x^{n+1}\in \mathcal{X}^{n + 1}$ , 
    $
    \mathrm{proj}_{\mathcal{X}^n} (\imath_{n+1}(g_n)x^{n+1})
    = 
    g_n(\mathrm{proj}_{\mathcal{X}^n} (x^{n+1})),
    $
    where $\mathrm{proj}_{\mathcal{X}^n}$ is the canonical projection map $\mathrm{proj}_{\mathcal{X}^n}: \mathcal{X}^{n+1}\to \mathcal{X}^{n}$.
    %
    \item\label{item:ass_iso} Let $n\geq 1$, $g_n \in G_n$, and $g_{n+1}\in G_{n+1}$. For $x^{n+1} = (x_1, \dots, x_{n+1})\in \mathcal{X}^{n+1}$, denote $(x^{n+1})_{n+1} = x_{n+1}$. Then, 
    $g_{n+1} = \imath_{n+1}(g_n)$
    if and only if, for all
    $
        x^{n+1} \in \mathcal{X}^{n+1},
        $ $
        (g_{n+1}x^{n+1})_{n+1}=x_{n+1}.
    $   
    \end{enumerate}
\end{definition}

Here, item~\ref{item:ordered_groups} gives an ordering of the sequence of groups by inclusion, \ref{item:sequential_action} ensures that this inclusion does not change the action of the groups on past data, and \ref{item:ass_iso} implies that the groups do not act on ``future'' data.
As a result, invariance of $X^{n-1}$ under $G_{n-1}$ is implied by invariance of $X^{n}$ under $G_{n}$ and the individual statements of invariance in~\eqref{eq:null} for each $n$ do not contradict each other.
The simplest example of a sequential action is when each $G_n=G_{n-1}\times H_n$ for some group $H_n$ and $G_n$ acts on $\mathcal{X}^n$ componentwise, i.e., $g_nX^n=(h_1X_1,\dots, h_nX_n)$. 
This setting is, among others, discussed in detail by~\citet{koning2023online}. A more complicated example of a sequential action (testing for sphericity) is given in Section~\ref{sec:orthogonal_ex}.

Under the assumption that the group action is sequential, we show that the null hypothesis of invariance is an online compression model.
They are models for computing online summaries, or compressed representations, of the observed data. 
When the data is generated by an online compression model, the techniques developed for conformal prediction can be used to construct a sequence of i.i.d. uniform statistics, which in turn give rise to a test martingale, as we discuss in Section~\ref{sec:conformal_martingales}. 
\citeauthor{vovk2005algorithmic} define online compression models in abstract terms; we use a simplified definition here. 
\begin{definition}[Online compression model]
\label{def:ocm}
An online compression model on $\mathcal{X}$ is a 3-tuple of sequences
$M=( (\sigma_n)_{n\in \mathbb{N}}, (F_n)_{n\in \mathbb{N}}, (Q_n)_{n\in \mathbb{N}}),$ 
where:
\begin{enumerate}
    \item  $(\sigma_n)_{n\in \mathbb{N}}$ is a sequence of statistics $\sigma_n=\sigma_n(X^n)$; we call $\sigma_n$ a summary of $X^n$,
    \item $(F_n)_{n\in\nats}$ is a sequence of functions such that $F_n(\sigma_{n-1}, X_n)=\sigma_n$,
    \item $(Q_n)_{n\in \nats}$ is a sequence of conditional distributions for $(\sigma_{n-1},X_n)$ given $\sigma_n$.
\end{enumerate}
\end{definition}

To show how sequential group invariance defines an online compression model, we first recall some group theory.
First, the orbit $G_n X^n$ of $X^n$ under the action of $G_n$ is the set of all values that are reached by the action of $G_n$ on $X^n$, i.e., $G_n X^n = \{gX^n: g\in G_n\}$. 
In order to identify each orbit, we pick a single element of $\mathcal{X}^n$ in each orbit---an orbit representative---and consider the map $\gamma_n:\mathcal{X}^n\to \mathcal{X}^n$ that takes each $X^n$ to its orbit representative.
We call $\gamma_n$ an orbit selector, and we assume that it is measurable. 
Such measurable orbit selectors are known to exist under weak regularity conditions on $\mathcal{X}^n$ and $G_n$~\citep[see][Theorem~2]{bondar1976borel} that hold in all the examples of this work.
Furthermore, because $G_n$ is a compact group, there exists a unique $G_n$-invariant probability distribution $\mu_n$, called the Haar measure~\citep[Chapter VII]{bourbaki2004integration}. 
The Haar measure plays the role of a uniform distribution on groups. 
Finally, it is a well-known fact that $X^n\mid \gamma_n(X^n) \eqind U\gamma_n(X^n)\mid \gamma_n(X^n)$, where $U\sim \mu_n$ independently of $X$~\cite[Theorem 4.4]{eaton1989group}. 

Together with the following proposition, these properties show that the sequential group invariance structure considered here defines an online compression model with $\sigma_n=\gamma_n(X^n)$.
\begin{proposition}\label{prop:update_rule}
    There exists a sequence $(F_n)_{n\in\mathbb{N}}$ of measurable functions $F_n:\mathcal{X}^{n-1}\times \mathcal{X} \to \mathcal{X}^{n}$ such that
    $F_n(\gamma_{n-1}(X^{n-1}), X_n)=\gamma_n(X^n)$, and
    $F_n( \ \cdot \ , X_n)$ is a one-to-one function of $\gamma_{n-1}(X^{n-1})$.
\end{proposition}

\begin{corollary}\label{cor:online_model}
 The tuple $((\gamma_n(X^n))_{n\in \mathbb{N}}, (F_n)_{n\in \mathbb{N}}, (\tilde{\mu}_n)_{n\in \mathbb{N}})$, where $\tilde{\mu}_n$ is the uniform distribution on $G_nX^n$ induced by the Haar measure $\mu_n$ on $G_n$, defines an online compression model on $\mathcal{X}$. 
\end{corollary}

\section{Testing group invariance with conformal martingales}
\label{sec:conformal_martingales}
The goal of this section is the construction of test martingales for the null hypothesis of distributional symmetry in~\eqref{eq:null} any time that a sequence of groups $(G_n)_{n\in \mathbb{N}}$ acts sequentially on the data $(X_n)_{n\in \mathbb{N}}$.  
To this end, the invariant structure of the null hypothesis $\mathcal{H}_0$ is used in tandem with conformal prediction to build a sequence of independent random variables $(R_n)_{n\in \mathbb{N}}$ with the following three properties: 
\begin{enumerate}
    \item The sequence $(R_n)_{n\in \mathbb{N}}$ is adapted to the data sequence with external randomization $(X_n,\theta_n)_{n\in \mathbb{N}}$, that is, for each $n\in\nats$, $R_n = R_n(X_n,\theta_n)$.
    \item Under any element of the null hypothesis $\mathcal{H}_0$ from~\eqref{eq:null}, $(R_n)_{n\in\mathbb{N}}$ is a sequence of independent and identically distributed $\mathrm{Uniform}([0,1])$ random variables. 
    \item The distribution of $(R_n)_{n\in\mathbb{N}}$ is not uniform when departures from symmetry are present in the data.
\end{enumerate}
The construction of these random variables is the subject of Section~\ref{sec:conformal_pred_invariance}---additional definitions are needed for their construction. 
In order to guide intuition,  Example~\ref{ex:sequential_ranks} shows a first example for testing exchangeability. 
Given their uniform distributions, the statistics $R_1, R_2, \dots$ have previously been called p-values~\citep[e.g.][]{vovk2005algorithmic,fedorova2012plugin}.
We opt against that terminology here, because typically only small p-values are interpreted evidence against the null hypothesis.
However, in this context, it is any deviation from uniformity that we interpret as evidence against the null hypothesis.

Once the sequence $(R_n)_{n\in\nats}$ has been built, test martingales against distributional invariance can be constructed. This is achieved by testing the uniformity of $(R_n)_{n\in\nats}$. Indeed, any time that $(f_n)_{n\in\mathbb{N}}$ is a sequence of functions $f_i:[0,1]\to \reals$ such that $\int f_i(r)\rmd r = 1$, the process $(M_n)_{n\in \nats}$ given by 
\begin{equation}\label{eq:test_mart}
    M_n := \prod_{i\leq n} f_i(R_i)
\end{equation}
is a test martingale for $\mathcal{H}_0$ with respect to $\filt$, where $\filt=(\sigma(R^n))_{n\in \nats}$.
This follows from the fact that $\expect{Q}{M_n\mid \sigma(R^{n-1})}=M_{n-1} \cdot \int f_n(r) \rmd r = M_{n-1}$, where we leverage independence and uniformity.
The functions $(f_n)_{n\in\nats}$ are known as calibrators~\citep{vovk2021values}. 
They can be taken to be any sequence of predictable estimators of the distribution of $R_1,R_2,\dots$~\citep{fedorova2012plugin}, so that the test martingale is expected to grow if the true distribution of the orbit ranks is not uniform, i.e., the null hypothesis is violated.  
The optimality of this procedure is discussed in Section~\ref{sec:mart_opt}.

\begin{example}[Sequential Ranks]\label{ex:sequential_ranks}
    Consider the case of testing exchangeability, that is, the case when  $\mathcal{X}=\mathbb{R}$  and each group $G_n$ is  $G_n = n!$, the group of permutations on $n$ elements.
    Consider, for each $n$, the random variables $\tilde{R}_n = \sum_{i \leq n}\indicator{X_i \leq X_n}$---the rank of $X_n$ among $X_1, \dots, X_n$. 
    It is a classic observation that each $\tilde{R}_n$ is uniformly distributed on $\{1, \dots, n\}$, and that $(\tilde{R}_n)_{n\in\mathbb{N}}$ is a sequence of independent random variables
     \citep{renyi1962extreme}. The random variables $\tilde{R}_1, \tilde{R}_2, \dots$ are called sequential ranks \citep{malov1996sequential}. 
     After rescaling and adding external randomization, a sequence of random variables $(R_n)_{n\in\mathbb{N}}$ can be built from $(\tilde{R}_n)_{n\in\mathbb{N}}$ such that $(R_n)_{n\in\mathbb{N}}$ satisfies items 1, 2 and 3 at the start of this section. 
    Furthermore, if we denote the uniform measure on $n!$ by $\mu_n$, then $\tilde{R}_n$ can also be obtained from 
    $
        n^{-1}\tilde{R}_n = \mu_n\{g: (gX_n)_n \leq X_n\}.
    $
    While this rewriting may seem esoteric at this point, it turns out to be the correct point of view for generalization as  there exists an analogue of the uniform probability distribution on every compact group---its Haar measure.  
\end{example}

\subsection{Conformal prediction under invariance}
\label{sec:conformal_pred_invariance}
In general, the statistics $R_n$ will be designed to measure how strange the observations $X^n$ are in contrast to what would be expected under distributional invariance. 
To this end, the values of $X^n$ are compared to those in the orbit of $X^n$ under the action of $G_n$.
In order to measure the ``strangeness'' of the observations in their orbit, we use an adaptation of the conformity measures introduced by~\citet{vovk2005algorithmic}. 
\begin{definition}[Conformity measure of invariance]
\label{def:nmi}
We say that $\alpha^n = \alpha^n(X^n)$ is a conformity measure of invariance at time $n$ if the following hold:
\begin{enumerate}[label=(\roman*)]
    \item\label{item:con_scores} $\alpha^{n}=(\alpha_1,\dots, \alpha_n)$, where $\alpha_i = A_n(X_i, \gamma_n(X^n))$ for a function $A_n:\mathcal{X}\times \mathcal{X}^n\to \reals$.
    \item\label{item:action_def} If  $\alpha^n(X^n)= \alpha^n(X'^n)$ for $X^n, X'^n\in \mathcal{X}^n$, then $\alpha^n(gX^n)=\alpha^n(gX'^n)$ for all $g\in G_n$.
\end{enumerate}
\end{definition} 
Item~\ref{item:action_def} in Definition~\ref{def:nmi}  is an addition to the definition by \citet{vovk2005algorithmic}. 
It ensures that the action of $G_n$ on $\mathcal{X}^n$ induces an action on the conformity measures, that is, it implies that the action of $G_n$ on $\alpha^n$ defined by $g\alpha^n := \alpha^n(gX^n)$ is well-defined.

The distribution of the conformity measures under the null hypothesis can be obtained by leveraging the distributional invariance. 
Indeed, as discussed in Section~\ref{sec:seq_action}, the distribution of $X^n$ conditional on $\gamma(X^n)$ is characterized by the Haar measure.
Similar to what happened in Example~\ref{ex:sequential_ranks}, this distribution can be used to rank the observed value of the conformity score $\alpha_n$ among all its possible values on the orbit of the data. 
This idea gives rise to the (smoothed) orbit ranks $(R_n)_{n\in \mathbb{N}}$ in the next definition.
\begin{definition}[Smoothed Orbit Ranks]
\label{def:orbit_ranks}
    Fix $n\in \nats$ and let $\alpha^n$ be a conformity measure of invariance at time $n$. We call $R_n$, defined by 
\begin{equation}\label{eq:pvals}
    R_n = \mu_n(\{g\in G_n: (g\alpha^n)_n < \alpha_n\})+\theta_n \mu_n(\{g\in G_n: (g\alpha^n)_n=\alpha_n\}),
\end{equation} 
a (smoothed) orbit rank, where $\mu_n$ denotes the Haar probability measure on $G_n$ and the sequence $\theta_1,\theta_2,\dots\stackrel{\mathrm{i.i.d}}{\sim} \mathrm{Uniform}[0,1]$ is independent of $\alpha^n$. 
\end{definition}

Note that, if the distribution of $\alpha^n$ conditional on $\gamma_n(X^n)$ is continuous, then the smoothing plays no role in~\eqref{eq:pvals}, and $R_n\perp \theta_n$.
Furthermore, if $A_n$ in Item~\ref{item:con_scores} of Definition~\ref{def:nmi} is chosen properly, a small orbit rank $R_n$ indicates that the observed value of $\alpha_n$ is strange (not conform) compared to the values it would have attained on different elements in the orbit of the data.
Alternatively, one can think of $R_n$ as the CDF of the distribution of $\alpha_n$ conditional on $\gamma_n(X^n)$ evaluated in the data (with added randomization).
It follows---and this is shown in Theorem~\ref{thm:iid_pvals}---that each $R_n$ is uniformly distributed on $[0,1]$.
\citet[Theorem 11.2]{vovk2005algorithmic} show that, if the data is generated by an online compression model, then $R_1,R_2,\dots$ are also independent. 
Since Corollary~\ref{cor:online_model} shows that a sequential group invariance structure defines an online compression model, it follows that the smoothed orbit ranks form an i.i.d. uniform sequence under the null hypothesis. 
This is stated in the next theorem, for which we provide a direct proof for completeness in~\ref{app:proof} in the supplementary material. 

\begin{theorem}\label{thm:iid_pvals}
    Suppose that the action of $(G_n)_{n\in \nats}$ on $(\mathcal{X}^n)_{n\in \nats}$ is sequential and that $(X_n)_{n\in \nats}$ is generated by an element of $\mathcal{H}_0$. 
    Then $R^n \perp \gamma_n(X^n)$ for each $n$ and the distribution of $(R_n)_{n\in\nats}$ is given by $U^\infty$. 
\end{theorem}

\section{Optimality}\label{sec:mart_opt}

In this section, we show that any martingale based on the smoothed orbit ranks as in~\eqref{eq:test_mart} can be thought of as likelihood ratio processes, and that 
they are log-optimal against the implicit alternative for which they are built.
Indeed, let $P$ be a distribution such that, conditionally on $R^{n-1}$, $R_n$ has density $f_n$ under $\tilde{P}$ for all $n$.
Here, we use $\tilde{P}$ to denote the distribution $P$ with added external randomization, i.e. $\tilde{P}:=P\times \mathcal{U}^\infty$, where $\mathcal{U}^\infty$ is the uniform distribution on $[0,1]^\infty$.
Analogously, for each $Q\in \mathcal{H}_0$, define $\tilde{Q}:=Q\times \mathcal{U}^\infty$.
Observe that $M_n=\prod_{i\leq n} f_i(R_i)$ equals the likelihood ratio between $\tilde{P}_{R^n}$ and $\tilde{Q}_{R^n}$, which equals the uniform distribution for any $Q\in \mathcal{H}_0$.
Surprisingly, $M_n$ is also the likelihood ratio of the full data between a distribution $P$ such that $R_n \perp \gamma_n(X^n)$ and an appropriately chosen distribution $Q^*\in\mathcal{H}_0$, as shown in the following proposition.
\begin{proposition}\label{prop:conf_is_lr} 
Suppose that $\alpha_n(X^n) = A_n( X_n , \gamma_n(X^n))$, where $A_n(\ \cdot \ , \gamma_n(X^n))$ is a one-to-one function for each $n\in \mathbb{N}$. Furthermore, suppose that $P$ is any distribution under which $R_n \perp \gamma_n(X^n)$ for each $n$. If $M_n = \prod_{i\leq n} f_i(R_i)$ and each $f_i$ is the conditional distribution of $R_i$ given $R^{i-1}$, then, for $Q\in \mathcal{H}_0$, 
\begin{equation}\label{eq:conf_is_lr}
     \tilde{Q}\left(M_n= \frac{\mathrm{d}P}{\mathrm{d}Q^*}(X^n)\right)=1,
\end{equation} 
where $Q^*$ denotes the distribution under which the marginal distribution of $\gamma_n(X^n)$ coincides with that under $P$, and such that $X^n\mid \gamma_n(X^n)\stackrel{\mathcal{D}}= U\gamma_n(X^n)\mid \gamma_n(X^n)$, where $U\sim \mu_n$ independently from $\gamma_n(X^n)$. 
\end{proposition}
The next theorem uses the representation in~\eqref{eq:conf_is_lr} to show the log-optimality of $(M_n)_{n\in \mathbb{N}}$.
Its proof is heavily inspired by~\citet[Theorem 12]{koolen2022log}. 
\begin{theorem}\label{thm:loavev}
    Assume that $\alpha_n(X^n) = A_n( X_n , \gamma_n(X^n))$, where $A_n(\ \cdot \ , \gamma_n(X^n))$ is one-to-one all $n\in \mathbb{N}$ and let $P$ be any distribution such that $X^n\mid \gamma_n(X^n)$ has full support.
    Denote $f_n$ for the density of $R_n\mid R^{n-1}$ under $\tilde{P}$. 
    Let $\tau$ be any stopping time and $(E_n)_{n\in\nats}$ any test martingale for $\mathcal{H}_0$, both with respect to $\filt$---the filtration generated by the smoothed ranks. Then, it holds that
    \begin{equation}\label{eq:rank_optim}
    \expect{\tilde{P}}{\ln M_\tau }=\expect{\tilde{P}}{\ln \prod_{i=1}^\tau f_i(R_i)}\geq  \expect{\tilde{P}}{\ln E_\tau}.
    \end{equation}
    Moreover, if $\tilde{P}$ is such that $R^n\perp \gamma_n(X^n)$ for all $n$, then for any test martingale $E'$ for $\mathcal{H}_0$ w.r.t. $(\sigma(X^n))_{n\in \mathbb{N}}$---the full-data filtration---, it also holds that
    \begin{equation}\label{eq:optim}
    \expect{\tilde{P}}{\ln M_\tau}
    \geq  \expect{\tilde{P}}{\ln E'_\tau}.
    \end{equation}
\end{theorem}

The additional assumption of independence between $R^n$ and $\gamma_n(X^n)$ is necessary for~\eqref{eq:optim} to hold:
if $\tilde{P}$ is a distribution under which  $R_1,\dots, R_n \not\perp \gamma_n(x^n)$, then the conformal martingale is not in general a likelihood ratio as in~\eqref{eq:conf_is_lr}. 
For the deterministic stopping time $\tau=n$, the log-optimal statistic is $S_n = \prod_{i=1}^n f_n(R_1,\dots,R_n\mid \gamma_n(X^n))$, as it can be written as a likelihood ratio~\citep[see also][]{grunwald2019safe,koning2023online}.
However, the sequence $(S_n)_{n\in\nats}$ does not necessarily give rise to an anytime-valid test.
Using tests based on the sequential ranks circumvents this issue for such alternatives.

The optimality of $M_n$ in Theorem~\ref{thm:loavev} is contingent on oracle knowledge of the true distributions $f_1,f_2,\dots$, which are unknown in practice.
To counter this, past data can be used sequentially to estimate the true density. This ideas has previously been applied for testing exchangeability~\citep{vovk2005algorithmic, fedorova2012plugin}.
More precisely, for each $n$, let $\hat f_n$ be an estimator of $f_n$ based on $R^{n-1}$, and consider the martingale defined by $\prod_{i=1}^n \hat f_i(R_i)$.
While this is suboptimal with respect to an oracle that knows the true density, there is limited loss asymptotically if $\hat{f}_i$ is a good estimator of $f_i$. 
In order to judge if an estimator is good for the task at hand, consider the difference in expected growth per outcome for fixed $n$, i.e., 
\begin{equation}\label{eq:redundancy}
\frac1n \sum_{i=1}^n \mathbb{E}_{\tilde{P}}[ \log f(R_i) - \log \hat f_i(R_i)]
= 
\frac1n \sum_{i=1}^n \mathbb{E}_{\tilde{P}} [\KL(f\|\hat f_i)]
,
\end{equation}
where $\KL(f\|\hat g) = \int_0^1 f(r) \log f(r)/g(r) \rmd r$ denotes the Kullback-Leibler divergence whenever $f$ is absolutely continuous with respect to $g$, and the expectation on the right-hand side  of \eqref{eq:redundancy} is over past data (on which $\hat{f}_i$ depends). 
Estimators under which \eqref{eq:redundancy} tends to zero are known to exist under weak conditions on the true density $f$~\citep{1997_opper_haussler_density,cesa2001worst,grunwald2019tight}. 
Under more stringent assumptions---for example, if the density $f$ belongs to an exponential family---sequential Bayesian-update-type algorithms are known to guarantee that \eqref{eq:redundancy} tends to 0 as $n\to\infty$~\citep{kotlowski2011maximum}.

\section{Applications and Extension}
\label{sec:applications}
In this section, we discuss applications and an extension of the theory developed above. 

\subsection{Modification for Independence Testing}\label{sec:indep_test}

We now propose a minor modification of the conformal martingales from the previous section that can be used to test for independence. Formally, fix $K\in \nats$ and suppose that at each time point $n\in\nats$, a $K$-dimensional vector $X_n=(X_{1,n},\dots, X_{K, n})\in \mathcal{X}^K$ is observed. 
We are interested in testing the null hypothesis that states that: (1) for each $k = 1, \dots, K$ and each $n$ the  vectors $(X_{k,1},\dots, X_{k,n})$ are $G_n$-invariant, and (2) $(X_{k,1},\dots, X_{k,n})\perp (X_{k',1},\dots, X_{k',n})$ for all $k\neq k'\in \{1,\dots,K\}$. 
Under this hypothesis, the data is invariant under the sequential action of $(\tilde{G}_n)_{n\in \mathbb{N}}$ given by $\tilde{G}_n=G_n^K$,
acting on $\mathcal{X}^{K\times n}$ rowwise.
That is, the first copy of the group acts on $(X_{1,1},\dots, X_{1,n})$, the second on $(X_{2,2},\dots, X_{2,n})$, etc. 
This action is sequential anytime that the action of $(G_n)_{n\in\nats}$ is sequential on each of the $K$ data streams.

Based on the discussion above, a first idea to test for invariance under $\tilde{G}_n$ is to create $K$ test martingales and combine them through multiplication.
More specifically, we can treat each of the sequences $(X_{k, n})_{n\in\mathbb{N}}$, $k\in\{1, \dots, K\}$ as a separate data stream and compute the corresponding statistics in~\eqref{eq:pvals}, leading to $K$ sequences of uniformly distributed random variables $(R_{k,n})_{n\in\mathbb{N}}$.
If, for all $n\in \nats$ and $k\in\{1,\dots,K\}$, $f_{k, n}$ is a density on $[0,1]$ then, by independence, the sequence $(M_n')_{n\in\mathbb{N}}$ defined by $M_n'=\prod_{i=1}^n \prod_{k=1}^K f_{k, i}(R_{k, i})$ is a martingale under the null hypothesis.
However, this martingale would not be able to detect alternatives under which the marginals are group invariant, but not independent.
This stems from the fact that it only uses that the marginals are uniform under the null, while in fact a stronger claim is true: for each $n$, the joint distribution of $R_{k, n}$, $k\in \{1,\dots,K\}$, is uniform on $[0,1]^K$.
As a result of this observation, one can choose any sequence of joint density (estimators) $f_1, f_2, \dots$ on $[0,1]^K$ and create a test martingale by considering $M_n=\prod_{i=1}^n f_i(R_{1, i},\dots,R_{K, i})$.

In the case that $K=2$ and $G_n=n!$, this is the procedure that was recently employed by \citet{henzi2023rank}.
They discuss a specific choice of $f_n$, a histogram density estimator, that is able to detect departures from independence consistently under the stronger assumption that data are i.i.d.
One of their key insights is that independence of the data streams not only implies joint uniformity of the sequential ranks in their setting, but the two are actually equivalent. 
This equivalence breaks down if one does not assume that $X_{k,1}, X_{k,2} \dots$ are i.i.d. for all $k$. 
Finding conditions under which the independence of the streams and the joint uniformity of the rank distributions are equivalent so that a histogram density estimator might reliably detect independence in the more general setting, is future work.

\subsection{The orthogonal group and linear models}
\label{sec:orthogonal_ex}
Consider testing whether the data we observe are drawn from a spherically symmetric distribution, i.e., $\mathcal{X}=\mathbb{R}$ and $G_n=\mathrm{O}(n)$, where $\mathrm{O}(n)$ is the orthogonal group in dimension $n$. 
Testing for spherical symmetry is equivalent to testing whether the data are generated by a zero-mean Gaussian distribution. 
This follows from the fact that any distribution on $\reals^\infty$ for which the marginal of the first $n$ coordinates is spherically symmetric for any $n$, can be written as a mixture of i.i.d. zero-mean Gaussian distributions~\citep[Proposition 4.4]{bernardo2009bayesian}.
It follows that any process that is a supermartingale under all zero-mean Gaussian distributions is also a supermartingale under spherical symmetry and vice-versa. 
This implies that, for the purpose of testing with supermartingales, the two  hypotheses are equivalent. We show how this fits in our setting, and deffer the application to regression to the Supplementary Material. 

We now check that testing spherical symmetry fits in our setting, i.e., that Definition~\ref{def:seq_groups} is fulfilled. 
Consider the inclusion of $\mathrm{O}(n)$ in  $\mathrm{O}(n+1)$ given by
\begin{equation*}
    \imath_{n + 1}(O_n) 
    = \begin{pmatrix}
        O_n & 0\\
        0 & 1
    \end{pmatrix}
\end{equation*} 
for each  $O_n\in \mathrm{O}(n)$. 
Using the canonical projections in $\reals^n$, Definition~\ref{def:seq_groups} is readily checked.
Since the data are real, $\alpha^n$ can be chosen to be the identity for all $n$, i.e., $\alpha^n(X^n)=X^n$.
An orbit selector is given by $\gamma_n(X^n)=\|X^n\|e_1$, where $e_1$ is the unit vector $e_1=(1,0,\dots,0)$.
For simplicity, we assume that the distribution of $X^n$ has a density with respect to the Lebesgue measure for each $n$, so that $R_n=\mu_n(\{O_n\in \mathrm{O}(n): (O_nX^n)_n < X_n\})$---no external randomization is needed.
Rather than thinking of $\mu_n$ as a measure on $O(n)$, one can think of it as the uniform measure on $S^{n-1}(\|X^n\|)$. 
This way, $R_n$ can be recognized to be the relative surface area of the hyper-spherical cap with co-latitude angle $\varphi_n = \pi-\cos^{-1} (X_n/\|X^n\|)$.
\cite{li2010concise} shows that an explicit expression for this area is given by
\begin{equation}\label{eq:hyperspher}
R_n = \begin{cases}
1-\frac12 I_{\sin^2(\pi-\varphi_n)}\left(\frac{n-1}{2}, \frac12\right) & \text{if } \varphi_n > \frac\pi2,\\
 \frac12 I_{\sin^2(\varphi_n)}\left(\frac{n-1}{2}, \frac12\right) & \text{else},
\end{cases}
\end{equation}
where $I_x(a,b)$ denotes the regularized beta function, $I_x(a,b) = \frac{B(x, a, b)}{B(1, a, b)}$ for $B(x, a, b) = \int_0^x t^{a - 1}(1 - t)^{b - 1}$ for $0\leq x\leq 1$. 

Note that $\varphi_n>\frac\pi 2$ if and only if $X_n>0$ and that $\sin^2(\varphi_n)=1-\frac{X_n^2}{\|X^n\|^2}$, so that \eqref{eq:hyperspher} equals the CDF of the t-distribution with $n-1$ degrees of freedom evaluated in $t=\sqrt{n-1} X_n  / \|X^{n-1}\|$.
If $X^n\sim \mathcal{N}(0,\sigma^2 I_n)$, then $t$ is the ratio of a normally distributed random variable and an independent chi-squared-distributed random variable.
Therefore, $t$ has a t-distribution with $n-1$ degrees of freedom, so that we essentially perform a type of sequential t-test. 

This example can be extended to testing for centered spherical symmetry, i.e., whether $X^n=\mu\mathbf{1}_n+\epsilon^n$, where $\mathbf{1}_n$ is the all-ones $n$-vector, $\mu\in\reals$ and $\epsilon^n$ is spherically symmetric for every $n\in \nats$.
By similar reasoning as above, this is equivalent to testing whether the data is i.i.d. Gaussian with any mean/variance. 
Even more, by considering different isotropy groups, one can also cover the case where the mean $\mu$ is not fixed, but depends on covariates. The techniques needed in that case are similar; we show them  in~\ref{sec:linear_models} of the supplementary material. 

\section{Discussion}
\label{sec:discussion}
We have discussed how the theory of conformal prediction can be applied to test for symmetry of infinite sequences of data. 
Here we discuss two topics. First, the relationship to noninvariant conformal martingales. 
Second, whether smoothing is necessary when defining orbit ranks. 

\subsection{Noninvariant conformal martingales}
Not all online compression models correspond to a compact-group invariant null hypothesis.
An interesting example of this phenomenon is when the data are i.i.d. and exponentially distributed.
This distribution is invariant under reflections in any $45^\circ$ line (not necessarily through the origin), but these reflections do not define a compact group and therefore do not fit the setting discussed in this article.
Nevertheless, the sum of data points is a sufficient statistic for the data, so this model can still be seen as an online compression model with the sum being the summary.
More work is needed to find out whether conformal martingales are log-optimal against certain alternatives in such settings.

\subsection{The need for smoothing}
In situations when, conditionally on the orbit selector $\gamma_n(X^n)$, the conformity measure $\alpha^n(X^n)$ has a continuous distribution, the smoothing plays no role in \eqref{def:orbit_ranks}. This is the case for the rotations discussed in Section~\ref{sec:orthogonal_ex}. 
In certain other scenarios, smoothing can be avoided as well. Indeed, one can always define nonsmoothed orbit ranks, in opposition to the smoothed ranks $R_n$ from Definition~\ref{def:orbit_ranks}, by
$
    \tilde{R}_n
    := 
    \mu_n(\{g\in G_n: (g\alpha^n)_n \leq \alpha_n\}).
$
Notice that this nonsmooth version satisfies $\tilde{R}_n\leq R_n$. 
For a particular choice of increasing densities $f_1, f_2, \dots, $ on $[0,1]$---in the sense that $u\mapsto f_i(u)$ is increasing---, we have that the process $\tilde{M}_n := \prod_{i = 1}^n f_i(\tilde{R}_i)$ is bounded from above by the conformal martingale $M_n = \prod_{i = 1}^n f_i(R_i)$. 
Such a choice of increasing $f_i$ is natural when high values of $R_i$ (or $\tilde{R}_i$) are associated with departures from the null hypothesis. 
Then, any sequential test based on an upper threshold on $\tilde{M}_n$ inherits the anytime-valid type-I error  guarantees of $M_n$---exactly because $\tilde{M}_n \leq M_n$. 
This was previously noted by~\citet{vovk2003testing}.
However, the process $\tilde{M}_n$ may not be a martingale itself. 
Instead, a test martingale can sometimes directly be associated to $\tilde{R}_n$. 
For instance, in the setting of Example~\ref{ex:sequential_ranks} (testing exchangeability), the distribution of $\tilde{R}_n$ under the null hypothesis is known---it is uniformly distributed on $\{1, \dots, n\}$.
Therefore, we can construct likelihood ratio processes for the sequence of nonsmoothed ranks. Even more, there are parametric alternatives under which the exact distributions of the nonsmoothed ranks can be computed. 
This is the case for Lehmann alternatives where, under the null, each $X_i$ is assumed to be sampled from some continuous distribution with c.d.f. $F_i(x) = F_0(x)$ for some fixed $F_0$; under the alternative, $F_i(x) = 1 - (1 - F_0(x))^{\theta_i}$ for some $\theta_i$. From Theorem~7.a.1 of \citet{savage_contributions_1956} the distribution of $\tilde{R}_i$ can be derived, so that the likelihood ratio process of $\tilde{R}_i$ can be used for testing, thus avoiding external randomization. 

\section{Acknowledgements}
We thank the attendees of the Seminar on Anytime-Valid Inference  ``E-readers'' at \emph{Centrum Wiskunde \& Informatica} in Amsterdam for their input and valuable insights. In particular, we are grateful to Peter Gr\"unwald for feedback on a first version of this article, and Nick Koning for fruitful discussions. 

\bibliographystyle{elsarticle-harv} 
\bibliography{references}


\newpage

\part*{Appendix}
\setcounter{page}{1}
\appendix 

Appendix for ``Anytime-Valid Tests of Group Invariance through Conformal Prediction'' by
Tyron Lardy and Muriel Felipe P\'erez-Ortiz.

\section{Proofs}\label{app:proof}

\subsection{Proof of Theorem~\ref{thm:iid_pvals}}
\begin{proof}[Theorem~\ref{thm:iid_pvals}]
The proof can be divided in two main steps: (1) to show that, conditionally on $\gamma_n(X^n)$, $R_n$ is uniformly distributed for each $n$ and (2) to show that $R_1,R_2,\dots$ are also independent.
The second step is completely analogous to the proof of Theorem 3 by~\cite{vovk2002line}.
For each $n$, define the $\sigma$-algebra $\mathcal{G}_n=\sigma(\gamma_n(X^n),X_{n+1},X_{n+2},\dots)$. Notice that $\mathcal{G}_n$ contains---among others---all $G_n$-invariant functions of $X^n$ because $\gamma_n$ is a maximally invariant function of $X^n$---any other $G_n$-invariant function of $X^n$ is a function of $\gamma_n(X^n)$. Let $g'\in G_n$ such that $\gamma_n(X^n)=g'X^n$, then we have that $\{g\in G_n: (g\alpha^n)_n<\alpha_n\}=\{g\in G_n: (\alpha^n(g\gamma_n(X^n)))_n<\alpha_n\}g'$.
By the invariance of $\mu_n$---it is the Haar probability measure---, it follows that \[\mu_n(\{g\in G_n: (g\alpha^n)_n<\alpha_n\})=\mu_n(\{g\in G_n: (\alpha^n(g\gamma_n(X^n)))_n<\alpha_n\}).\]
An analogous identity can be derived for the second term in~\eqref{eq:pvals}. 
We have $\alpha_n\mid \mathcal{G}_n\stackrel{\mathcal{D}}{=} (\alpha^n(U\gamma_n(X^n)))_n\mid \mathcal{G}_n$. 

We will denote $F(b):=\mu(\{g\in G_n: (g\alpha^n)_n <b\})$ and define $G (\delta) = \sup\{b\in \mathbb{R}: F(b)\leq \delta\}$.
If $\alpha_n\mid \mathcal{G}_n$ is continuous, then $F$ is the CDF of that distribution, otherwise it is the CDF minus the probability of equality. 
In any case, $F$ is is nonincreasing and left-continuous. 
For any $\delta\in (0,1)$, we have that $F(G(\delta))=\delta'$ for some $\delta' \leq \delta$, with equality if $F$ is continuous in $G(\delta)$.
Then we can write
\begin{equation}\label{eq:cdf_decomp}
    \mathbb{P} (R_n \leq \delta \mid \mathcal{G}_n)= \mathbb{P}(R_n\leq \delta'\mid\mathcal{G}_n)+\mathbb{P}(\delta' < R_n \leq \delta\mid \mathcal{G}_n).
\end{equation} 
For any $\theta\in (0,1]$, we have that $R_n=F(\alpha_n)+ \theta (F(\alpha_n^{+}) -F(\alpha_n))\leq\delta'$ if and only if either $F(\alpha_n)<\delta'$ or $F(\alpha_n^{+}) -F(\alpha_n)=0$, which happens precisely when  $\alpha_n<G(\delta)$.
We therefore see
\[ \mathbb{P} (R_n \leq \delta' \mid \mathcal{G}_n)=\mathbb{P}(\alpha_n < G(\delta') \mid \mathcal{G}_n)= F(G(\delta'))=\delta'.\]
If $F$ is continuous in $G(\delta)$, then this shows that $\mathbb{P} (R_n \leq \delta \mid \mathcal{G}_n)=\delta$, since $\delta'=\delta$ in that case.
If $F$ is not continuous in $G(\delta)$, then we have that
\begin{align*}
    \mathbb{P}(\delta' < R_n \leq \delta\mid \mathcal{G}_n)&=\mathbb{P}(\delta' < F(\alpha_n)+\theta (F(\alpha^{+}_n)- F(\alpha_n))\leq \delta\mid \mathcal{G}_n).
\end{align*}
Notice that $\delta' < F(\alpha_n)+\theta ({F}(\alpha^{+}_n)- F(\alpha_n))\leq \delta$ if and only if $\alpha_n=G(\delta)$ and $\theta <(\delta-\delta')/(F(\alpha^{+}_n)- F(\alpha_n))$, so that we can write
\begin{align*}
    \mathbb{P}(\delta' < R_n \leq \delta\mid \mathcal{G}_n)
    &=\mathbb{P}(\alpha_n=G(\delta)\mid \mathcal{G}_n)\mathbb{P}\left(\theta \leq  \frac{\delta -\delta'}{F(G(\delta')^+)- F(G(\delta'))}\mid \mathcal{G}_n\right)\\
    &=({F}(G(\delta')^+)- {F}(G(\delta'))) \frac{\delta -\delta'}{({F}(G(\delta')^+)- {F}(G(\delta')))}\\
    &=\delta-\delta'.
\end{align*}
Putting everything together, we see that $\mathbb{P} (R_n \leq \delta \mid \mathcal{G}_n)=\delta$. This shows the first part, that $R_n$ has a conditional uniform distribution on $[0,1]$. 

For the second part of the proof, we show that the sequence $R_1, R_2, \dots$ is also an independent sequence. We have that $R_n$ is $\mathcal{G}_{n-1}$-measurable because it is invariant under transformations of the form $X^n \mapsto (gX^{n-1},X_n)$ for $g\in G_{n-1}$ \citep[see also][Lemma 2]{vovk2004universal}.
We proceed (implicitly) by  induction:
\begin{align*}
    \mathbb{P}(R_n\leq \delta_n,\dots,R_1\leq \delta_1\mid \mathcal{G}_n)&= \expect{}{\indicator{R_n\leq \delta_n,\dots,R_1\leq \delta_1}\mid \mathcal{G}_n}\\
    &=\expect{}{\expect{}{\indicator{R_n\leq \delta_n,\dots,R_1\leq \delta_1}\mid \mathcal{G}_{n-1}}\mid \mathcal{G}_n}\\
    &=\expect{}{\indicator{R_n\leq \delta_n}\expect{}{\indicator{p_{n-1}\leq \delta_{n-1},\dots,R_1\leq \delta_1}\mid \mathcal{G}_{n-1}}\mid \mathcal{G}_n}\\
    &=\expect{}{\indicator{R_n\leq \delta_n}}\delta_{n-1}\cdots \delta_1\\
    &=\delta_n\cdots \delta_1.
\end{align*}
It follows by the law of total expectation that 
\[ \mathbb{P}(R_n\leq \delta_n,\dots,R_1\leq \delta_1) = \delta_n\cdots \delta_1,\]
which shows that $R_1,R_2,\dots,R_n$ are independent and uniformly distributed on $[0,1]$ for any $n\in \mathbb{N}$.
This implies that the distribution of $R_1,R_2,\dots$ coincides with $U^\infty$ by Kolmogorov's extension theorem~\citep[see e.g.][Theorem~II.3.3]{shiryaev2016probability}. This shows the claim of the theorem. 
\end{proof}

\subsection{Proof of Proposition~\ref{prop:update_rule}}

\begin{proof}[Proposition~\ref{prop:update_rule}]
For $X^{n-1}\in \mathcal{X}^{n-1}$ and $X_n\in\mathcal{X}$, let $F_n(X^{n-1}, X_n) = \gamma_n((X^{n-1}, X_n))$, where, by a slight abuse of notation, we refer by $(X^{n-1}, X_n)$ to the concatenation of $X^{n-1}$ and $X_n$. We will show that $F_n$ has the claimed properties. First, we will show that the vectors $(\gamma_{n-1}(X^{n-1}), X_n)$ and $X^n$ are in the same orbit, so that also $\gamma_n( (\gamma_{n-1}(X^{n-1}), X_n))=\gamma_n(X^n)$.
    To this end, let $g'\in G_{n-1}$ denote the group element such that $g' X^{n-1}=\gamma_{n-1}(X^{n-1})$. Then it holds that
    \begin{align*}
        \{g(\gamma_{n-1}(X^{n-1}),X_n):g\in G_n\}&=\{g(g'X^{n-1}, X_n):g\in G_n\}\\
        &= \{ g \imath_{n}(g')X^n:g\in G_n\}\\
        &= \{ gX^n: g\in G_n\},
    \end{align*}
    where we called $X^n$ the concatenation of $X^{n-1}$ and $X_n$. This shows the first claim. For the second claim, that $F_n( \ \cdot \ , X_n)$ is one-to-one for each fixed $X_n$, we show that we can reconstruct $\gamma_{n-1}(X^{n-1})$ from $X_n$ and $\gamma_n(X^n)$.
    
    Pick any $g_{X_n}\in G_n$ such that $(g_{X_n}\gamma_n(X^n))_n=X_n$. 
    We furthermore know that there exists some $g\in G_n$ such that $gX^n=\gamma_n(X^n)$. 
    Note that $g_{X_n}g$ does nothing to the final coordinate of $X^n$, so by Assumption~\ref{def:seq_groups} there is a $g_{n-1}^*\in G_{n-1}$ such that $g_{X_n}gX^n=\imath(g_{n-1}^*)X^n$. 
    Then we see
    \begin{align*}
        \{\imath(g_{n-1}) g_{X^n}\gamma_n(X^n): g_{n-1}\in G_{n-1}\}&= \{\imath(g_{n-1})  g_{X^n}gX^n : g_{n-1}\in G_{n-1}\}\\
        &= \{ \imath(g_{n-1}) \imath(g_{n-1}^*) X^n: g_{n-1}\in G_{n-1}\}\\
        &= \{ \imath(g_{n-1})X^n: g_{n-1}\in G_{n-1}\}.
    \end{align*}
    We find that $G_{n-1} \mathrm{proj}_{n-1}( g_{X_n}\gamma_n(X^n))= G_{n-1}X^{n-1}$ and therefore $\gamma_{n-1}(\mathrm{proj}_{n-1}(g_{X_n}\gamma_n(X^n)))=\gamma_{n-1}(X^{n-1})$.
\end{proof}

\subsection{Proof of Proposition~\ref{prop:conf_is_lr}}

The proof of Proposition~\ref{prop:conf_is_lr} follows directly from Lemma~\ref{lem:one-to-one}. It states that, with probability one, enough of the original data can be recovered using the smoothed ranks and the orbit representative. We state Lemma~\ref{lem:one-to-one}, prove Proposition~\ref{prop:conf_is_lr} and then prove Lemma~\ref{lem:one-to-one}.

\begin{lemma}\label{lem:one-to-one}
Suppose, for each $n\in \mathbb{N}$, that $\alpha_n( \ \cdot \ , \gamma_n(X^n))$ is a one-to-one function of $X_n$, then 
there exists a map $D_n:[0,1]^n\times \mathcal{X}^n\to  [0,1]^n\times \mathcal{X}^n$ s.t. for any $Q\in \mathcal{H}_0$
$\tilde{Q}(D_n(R^n, \gamma_n(X^n))= (\tilde{\theta}^n,X^n))=1.$
Here, $\tilde{\theta}^n = (\tilde{\theta}_n)_{n\in\nats} $ is the sequence given by $\tilde{\theta}_n = \theta_n \indicator{\mu_n(\{g\in G_n : (g\alpha^n)_n = \alpha_n\}) \neq 0}$.
\end{lemma}

\begin{proof}[Lemma~\ref{prop:conf_is_lr}]
    Consider, without loss of generality, the case that $\alpha_n(X^n) = X_n$. Because of the independence of $R_n$ and $\gamma_n$ under $P$ and the assumption that the marginal distribution of $\gamma_n$ under $Q^*$ and under $P$ are equal, $M_n = \frac{\rmd \tilde{P}(R^n, \gamma_n(X^n))}{\rmd \tilde{Q}^*(R^n, \gamma_n(X^n))}$. Using the sequence of functions $(D_n)_{n\in\nats}$ from Lemma~\ref{lem:one-to-one} and that the external randomization is independent of $X^n$, the claim follows.
\end{proof}

\begin{proof}[Lemma~\ref{lem:one-to-one}]
As in the proof of Theorem~\ref{thm:iid_pvals}, we will denote $F(b)=\mu_n(\{g\in G_n: (g\alpha^n)_n <b\})$ and define $G(\delta)=\sup\{b\in \mathbb{R}: F(b)\leq \delta\}$.
Furthermore, we will write $\mathbb{P}_{\alpha_n\mid \gamma_n(X^n)}$ for the distribution of $\alpha_n$ given $\gamma_n(X^n)$ and denote its support by 
\[\mathrm{supp}(\mathbb{P}_{\alpha_n\mid \gamma_n(X^n)}):=\{x\in \mathbb{R}\mid \text{for all } I \text{ open, if } x\in I \text{ then } \mathbb{P}_{\alpha_n\mid \gamma_n(X^n)}(I)>0\},\]
If $b\in \mathrm{int}(\mathrm{supp}(\mathbb{P}_{\alpha_n\mid \gamma_n(X^n)}))$, then there exists an open interval $B$ with $b\in B$ and $B\subseteq \mathrm{supp}(\mathbb{P}_{\alpha_n\mid \gamma_n(X^n)}))$.
For all $c\in B$ with $c>b$, we have that $F(c)-F(b)= \mathbb{P}_{\alpha_n\mid \gamma_n(X^n)} ([b,c))>0$, since $[b,c)$ contains an open neighborhood of an interior point of the support. 
It follows that $F(c)>F(b)$. In words, there are no points $c$ to the right of $b$ such that $F(c) > F(b)$. Consequently, we have
\[G(F(b))=\sup\{a\in \mathbb{R}: F(a)\leq F(b)\}=b.\]
In a similar fashion, we can conclude that the same identity holds if $b\in \mathrm{supp}(\mathbb{P}_{\alpha_n\mid \gamma_n(X^n)})\setminus\mathrm{int}(\mathrm{supp}(\mathbb{P}_{\alpha_n\mid \gamma_n(X^n)}))$.
Notice furthermore that  $G(R_n)=G(F(\alpha_n)+\theta_n({F}(\alpha_n^+)-F(\alpha_n)))=G(F(\alpha_n))$ whenever $\theta_n<1$, which happens with probability one.
Together with the fact that $\mathbb{P}_{\alpha_n\mid \gamma_n(X^n)} (\mathrm{supp}(\mathbb{P}_{\alpha_n\mid \gamma_n(X^n)}))=1$, this gives $\mathbb{P}_{\alpha_n\mid \gamma_n(X^n)}(G(R_n)=\alpha_n)=1$, so also $\mathbb{P}(G(R_n)=\alpha_n)=1$. 
If $({F}(G(R_n)^+)-{F}(G(R_n)))=\mu_n(\{g\in G_n : (g\alpha^n)_n = \alpha_n\}) = 0$, set $\tilde{\theta}_n=0$.
If $\mu_n(\{g\in G_n : (g\alpha^n)_n = \alpha_n\}) > 0$, then it follows that $\mathbb{P}(\theta_n=(R_n-{F}(G(R_n)))/ ({F}(G(R_n)^+)-{F}(G(R_n))))=1$, so set $\tilde{\theta}_n =(R_n-{F}(G(R_n)))/ ({F}(G(R_n)^+)-{F}(G(R_n))) $.
Since $\alpha_n(\cdot, \gamma_n(X^n))$ is one-to-one by assumption, its inverse maps $\alpha_n$ to $X_n$. 
By Proposition~\ref{prop:update_rule}, there also exists a map from $X_n$ and $\gamma_n(X^n)$ to $\gamma_{n-1}(X^{n-1})$. 
At this point, we can repeat the procedure above to recover $X_{n-1}$ from $(R_{n-1}, \gamma_{n-1}(X^{n-1}))$, from which we can then recover $\gamma_{n-2}(X^{n-2})$, etc.
Together, all of the maps involved give the function as in the statement of the proposition.
\end{proof}

\subsection{Proof of Theorem~\ref{thm:loavev}}
\begin{proof}[Theorem~\ref{thm:loavev}]
We first show~\eqref{eq:optim}. Assume that $\tilde{P}$ is such that $R^n\perp \gamma_n(X^n)$ for all $n$.
Let $Q^*$ denote the distribution under which the marginal of $\gamma_n(X^n)$ coincides with that under $P$, and such that $X^n\mid \gamma_n(X^n)\stackrel{\mathcal{D}}= U\gamma_n(X^n)\mid \gamma_n(X^n)$, where $U\sim \mu_n$ is uniform on $G_n$ and independent from $\gamma_n(X^n)$.
First note that
\begin{align*}
    \tilde{Q}^*\left(\prod_{i=1}^\tau f_i(R_i) = \frac{\mathrm{d} {P}}{\mathrm{d} {Q}^*} (X^\tau)\right) &\geq \tilde{Q}^*\left(\forall t: \prod_{i=1}^t f_i(R_i) = \frac{\mathrm{d} P}{\mathrm{d} Q^*} (X^t)\right) \\
    &=1- \tilde{Q}^*\left(\exists t: \prod_{i=1}^t f_i(R_i) \neq \frac{\mathrm{d} {P}}{\mathrm{d} Q^*} (X^t)\right)\\
    &= 1-\tilde{Q}^*\left(\bigcup_{t=1}^\infty \left\{\prod_{i=1}^t f_i(R_i) \neq \frac{\mathrm{d} P}{\mathrm{d} {Q}^*} (X^t)\right\}\right)\\
    &\geq 1- \sum_{t=1}^\infty \tilde{Q}^*\left(\left\{\prod_{i=1}^t f_i(R_i) \neq \frac{\mathrm{d} P}{\mathrm{d} Q^*} (X^t)\right\}\right)=1. 
\end{align*}
In the last inequality, we used Lemma~\ref{lem:one-to-one}. Since the distribution of $X\mid \gamma_n(X^n)$ has full support under $P$, we have that $\tilde{P}\ll \tilde{Q}^*$, so it  also holds that $\tilde{P}\left(\prod_{i=1}^\tau f_i(R_i) = \frac{\mathrm{d} {P}}{\mathrm{d} {Q}^*} (X^\tau)\right)=1$. We have shown that $M_{\tau}$ is a modification of the likelihood ratio evaluated at $X^{\tau}$. We now show that the latter is optimal. 

Denote $\ell_n=\frac{\mathrm{d}P}{\mathrm{d}Q^*}(X^n)$ and let $f(\alpha)=\expect{\tilde{P}}{\ln((1-\alpha)\ell_\tau +\alpha E'_\tau)}$; a concave function. We will show that the derivative of $f$ in $0$ is negative, which implies that $f$ attains its maximum in $\alpha =0$. This in turn implies our claim. Indeed,
\begin{align*}
    f'(0)&= \expect{\tilde{P}} {\frac{E'_\tau-\ell_\tau}{\ell_\tau} }\\
    &= \sum_{i=1}^\infty \expect{\tilde{P}}{ \frac{E'_i}{\ell_i} \indicator{\tau=i}}-1\\
    &= \sum_{i=1}^\infty \expect{\tilde{Q}^*} {E'_i \indicator{\tau=i}}-1\\
    &= \expect{\tilde{Q}^*}{E'_\tau}-1\leq 0,
\end{align*}
where we use that differentiation and integration can be interchanged, because
\[|f'(\alpha)|=\left| \frac{E'_\tau - \ell_\tau}{(1-\alpha)\ell_\tau + \alpha E'_\tau} \right| \leq \max\left\{ \frac{1}{1-\alpha}, \frac1\alpha\right\},\]
so that the dominated convergence theorem is applicable.
Finally, this gives that $\expect{\tilde{P}}{\ln \prod_{i=1}^\tau f(R_i)}= \expect{\tilde{P}}{\ln E'_\tau} \geq \expect{\tilde{P}}{\ln E'_\tau}$.
The proof of~\eqref{eq:rank_optim} follows from the same argument, but using $\ell'_n=\frac{\mathrm{d}P}{\mathrm{d}Q^*}(R^n)$.
\end{proof}

\newpage

\section{Linear models and isotropy groups}
\label{sec:linear_models}
The rotational symmetry described in Section~\ref{sec:orthogonal_ex} is that of symmetry around the origin, which we argued is equivalent to testing whether $X_i\sim \mathcal{N}(0,\sigma)$ for some $\sigma\in \mathbb{R}^+$. 
Of course, there are many applications where it is not reasonable to assume that the data is zero-mean and it is more interesting to test whether the data is spherically symmetric around some point other than the origin.
One particular instance of such noncentered sphericity is to test whether, for each $n$, the data can be written as $X^n=\mu\mathbf{1}_n + \epsilon^n$, where $\mu \in \mathbb{R}$, the error $\epsilon^n$ is spherically symmetric and $\mathbf{1}_n$ is the $n$-vector of all ones. 
If $\mu$ is known, we can test for spherical symmetry of $X^n-\mu\mathbf{1}_n$ under $\mathrm{O}(n)$ and the problem reduces to that of the previous section. 
It is still possible treat the more realistic case where $\mu$ is unknown because the null model is still symmetric under a family of rotations. Notice the following: for any $O_n\in \mathrm{O}(n)$ it holds that $O_nX^n= \mu O_n\mathbf{1}_n + O_n\epsilon^n$. 
Unless $\mu=0$, it follows that $X^n\stackrel{\mathcal{D}}= O_nX^n$ every time that $O_n \mathbf{1}_n=\mathbf{1}_n$.
That is, the null distribution of $X^n$ is invariant under the isotropy group of $\mathbf{1}_n$, i.e. $G_n=\{O_n\in \mathrm{O}(n): O_n\mathbf{1}_n=\mathbf{1}_n\}$. 
Invariance under the action of $G_n$ has previously appeared in the literature as centered spherical symmetry~\citep{smith1981random}. 
Through the lens of test martingales, testing sequentially for centered spherical symmetry is equivalent to testing whether the data was generated by any Gaussian.
This holds because any probability distribution on $\mathbb{R}^\infty$ for which the marginal of the first $n$ coordinates is centered spherically symmetric for any $n$ can be written as a mixture of Gaussians~\cite[Theorem 8.13]{smith1981random,eaton1989group}. 

Using some geometry, a test is readily obtained. Note that we can write $X^n = X^n_{\mathbf{1}_n} + X^n_{\perp \mathbf{1}_n}$, where $X^n_{\mathbf{1}_n} =  \frac{\langle X^n, \mathbf{1} \rangle}{n}\mathbf{1}_n$ is the projection of $X^n$ onto the span of $\mathbf{1}_n$, and $X^n_{\perp \mathbf{1}_n}$ the projection onto its orthogonal complement. We have that $g X^n= X^n_{\mathbf{1}_n} + gX^n_{\perp \mathbf{1}_n}$ for any $g\in G_n$.
Consequently, the orbit of $X^n$ under $G_n$ is given by the intersection of $S^{n-1}(\|X^n\|)$ and the hyperplane $H_n(X^n)$ defined by $H_n(X^n)=\{x'^n\in \mathbb{R}^n: \langle x'^n, \mathbf{1}_n\rangle=\langle X^n, \mathbf{1}_n\rangle\}$.
There is a unique line that is perpendicular to $H_n(X^n)$ and passes through the origin $0_n=(0,\dots,0)$; it intersects $H_n(X^n)$ in the point $0_{H_n}:=\frac{\langle X^n, \mathbf{1}_n\rangle}{n}\mathbf{1}_n$.
For any $x'^n\in S^{n-1}(\|X^n\|)\cap H_n(X^n)$, Pythagoras' theorem gives that $\|x'^n-0_{H_n}\|^2=\|X^n\|^2-\|0_{H_n}-0_n\|^2$.
In other words, $S^{n-1}(\|X^n\|)\cap H_n(X^n)$ forms an $(n-2)$-dimensional sphere of radius $(\|X^n\|^2-\|0_{H_n}-0_n\|^2)^{1/2}$ around $0_{H_n}$.
If one considers the projection of this sphere on the $n$-th coordinate, then the highest possible value is given by $\|X^n\|$, and the lowest value therefore by $\frac{\langle X^n, \mathbf{1}_n\rangle}{n}-\frac12 (\|X^n\|-\frac{\langle X^n, \mathbf{1}_n\rangle}{n})$.
The relative value of $X_n$ is therefore given by $\tilde{X}_n:=X_n-\frac{\langle X^n, \mathbf{1}_n\rangle}{n}+\frac12 (\|X^n\|-\frac{\langle X^n, \mathbf{1}_n\rangle}{n})$.
As a result, $R_n$ is the relative surface area of the $(n-2)$-dimensional hyper-spherical cap with co-latitude angle $\varphi=\pi-\cos^{-1}( \tilde{X}_n/(\|X^n\|^2-\|0_{H_n}-0_n\|^2)^{1/2})$, so that equation~\eqref{eq:hyperspher} can again be used to determine $R_n$.
With this construction, we recover what~\cite{vovk2023power} refers to as the ``full Gaussian model'', which is an online compression model that is defined in terms of the summary statistic $\sigma_n=(\langle X^n, \mathbf{1}_n\rangle, \|X^n\|)$.

This model can be extended to the case in which there are covariates, i.e. $X_n=(Y_n, Z^d_n)$ for some $Y_n\in \mathbb{R}$ and $Z^d_n\in \mathbb{R}^d$.
Denote $Z_n$ for the matrix with row-vectors $Z^d_n$ and, as is a standard assumption in regression, assume that $Z_n$ is full rank for every $n$.
The model of interest is $Y^n=  Z_n  \beta + \epsilon^n$ where $\beta\in \mathbb{R}^d$ and $\epsilon^n$ is spherically symmetric for each $n$.
Similar to the reasoning above, this model is invariant under the intersection of the isotropy groups of the column vectors of $Z_n$, i.e. $G_n=\{O_n\in \mathrm{O}(n): O_nZ_n=Z_n\}$.
The orbit of $X^n$ under $G_n$ is given by the intersection of $S^{n-1}(\|X^n\|)$ with the intersection of the $d$ hyperplanes defined by the columns of $Z_n$, so that for $\alpha^n(Y^n,Z_n)= Y^n$, computing $R_n$ is analogous.
Interestingly, however, it does not always hold that testing for invariance under $G_n$ is equivalent to testing for normality with mean $Z_n \beta^d$.
A sufficient condition for the equivalence to hold is that $\lim_{n\to \infty}(Z_n'Z_n)^{-1}=0$, which is essentially the condition that the parameter vector $\beta$ can be  consistently estimated by means of least squares~\cite[Section 9.3]{eaton1989group}. 

\end{document}